\documentstyle[aps,prl,twocolumn]{revtex}

\begin{document}
\draft


\title{Observation of the Metal-Insulator Transition in Two-Dimensional 
n-type GaAs}

\author{Y. Hanein$^{1,2}$, D. Shahar$^{1,2}$, J. 
Yoon$^{2}$, C.C. Li$^{2}$, D.C. Tsui$^{2}$ and Hadas Shtrikman$^{1}$}

\address{$^{1}$ Dept. of Condensed Matter Physics, Weizmann Institute, 
Rehovot 76100, Israel}

\address{$^{2}$ Dept. of Electrical Engineering, Princeton University, 
Princeton, New Jersey 08544}
\maketitle

\begin{abstract}

The observation of a carrier-density driven metal-insulator transition 
in n-type GaAs-based heterostructure is reported. Although 
weaker than in comparable-quality p-type GaAs samples, the main features 
of the transition are rather similar. 
\end{abstract}



The recent discovery\cite{SVKrav94}  of a 
metal-insulator transition (MIT) in two-dimensional (2D) Si-based 
samples has been met with a certain degree of skepticism. 
This 
skepticism was largely founded on several successful theoretical 
studies \cite{Abrahams1979}, which enjoyed a strong support from 
experiments 
\cite{Bishop1980,Uren1981}. 
Therefore, it was only when the pioneering work of Kravchenko 
and his collaborators was followed 
by several other groups investigating various other materials 
\cite{Pudalov1997,DPopovic97,Ismail,Colridge,Hanein1998,Simmons1998a,Simmons1998b,SPapadakis1998} 
that the initial 
objections began to dissolve, and the MIT in 2D has become a mainstream 
field of study, where old \cite{SFinkel84}
and new 
\cite{VDobrosavljevic97,DBelitz97,PPhillips98,SHe97,Varma98,Castellani,Chakravarty,Altshuler98}
theoretical approaches are discussed.

Conspicuously missing from the list of semiconducting materials 
exhibiting the MIT in 2D has been the n-type GaAs system. In fact, in a 
recent study \cite{Dahm} of a high-density n-type GaAs/AlGaAs system, the 
MIT was 
clearly shown to be absent down to milli-Kelvin temperatures. The 
absence of the MIT in that sample was attributed to the 
relatively low level of 
the electron-electron interaction strength. 
This is because the low mass of the electrons in GaAs, which is a 
contributing factor in many of the unique phenomena associated with 2D 
systems, reduces the relative strength of these interactions. 
It is convenient to express the relative interaction-strength in terms of 
$r_{s}=E_{c}/E_{F}$, the ratio of the typical electron-electron 
interaction 
energy, $E_{c}=e^{2}/\epsilon r$, to the Fermi energy, $E_{F}$ ($e$ is 
the elementary charge, $\epsilon$ the dielectric constant and $r$ is 
the average distance between carriers). The lower mass increases 
$E_{F}$ (for a given carrier density) and therefore low carrier-mass 
systems are usually characterized by small $r_{s}$ values.
With the accumulating evidence for the existence of the MIT in several 
2D systems it became clear that a high value of $r_{s}$, usually $>5$,
is a prerequisite for the observation of the MIT. In Ref. \cite{Dahm} 
the system turns (strongly) insulating, presumably due to disorder, 
at a density $n>1.4\cdot 10^{11}$ cm$^{-2}$, 
corresponding to $r_{s}>1.4$, which is clearly
too low for the observation of the MIT.

To check these conjectures, and to facilitate a comparison with our 
p-type samples, we grew and measured a low-disorder n-type 
GaAs/AlGaAs sample.  
The sample was grown by a molecular beam epitaxy (MBE) technique on 
semi-insulating (311)A  substrate, which is
similar to that used for our p-type samples.
An appealing advantage of 
the (311)A growth 
orientation in GaAs/AlGaAs structures is associated with the 
amphoteric nature of Si on this growth plane \cite{Pavesi95}. By 
adjusting the 
growth parameters one can obtain either n-type or p-type material: 
At high growth 
temperature ($T\geq 600$ deg C) and low impinging arsenic flux (As/Ga 
flux ratio of 1/5), the Si will occupy the As sites and turn into a 
p-type dopant, and at low growth temperature ($T\leq 500$ deg C) 
and high impinging arsenic flux (As/Ga 
flux ratio $>10$) the Si will be forced to occupy the Ga sites and 
the material will be n-type. 
An added advantage of MBE growth on the (311)A plane is a reduction 
in background impurity incorporation, particularly carbon 
\cite{Wang}, in comparison with growth on the more common (100) plane.
For the (n-type) sample in this work our growth comprises a GaAs 
buffer, 
followed by a $500$\AA~  AlGaAs spacer layer, $500$\AA~  Si-doped 
n-AlGaAs, $150$\AA~  un-intentionally doped AlGaAs capped with 
$150$\AA~  GaAs layer. 
Incorporating a large spacer layer in the growth resulted in a high peak 
mobility of $2.2\cdot 10^{6}$ cm$^2$/Vsec at 
$n=13.6\cdot 10^{10}$ cm$^{-2}$.  
The measurements in this study were performed using a Hall-bar 
geometry with current flowing in the
[$\bar{2}33$] direction.
A Ti/Au gate was deposited on the top of the sample and was used to 
control the carrier density, with an effective range of 
$n=1.2-13.6\cdot 10^{10}$ cm$^{-2}$. 

As mentioned earlier, mounting 
evidence indicate that the observation of the MIT in 2D requires 
high-$r_{s}$ samples. While the route taken in most previous studies 
centered around obtaining high-$r_{s}$ by using high carrier-mass systems, 
we utilized the high quality of our sample to attain low 
densities, while still maintaining a conducting state. 
Our ability to reach such low density enabled large values of 
$r_{s}$, $r_{s}\leq 4.8$,
to be conveniently studied.

The high quality of our sample is attested to in Fig. 1, where we plot 
magnetic field traces of the longitudinal ($\rho_{xx}$) and 
Hall ($\rho_{xy}$) components of the resistivity tensor, obtained 
from sample H678D at $T=35$ and $38$ mK. 
Two densities are shown, $n=1.3\cdot 10^{10}$ 
cm$^{-2}$ in Fig. 1a and $n=6\cdot 10^{10}$ 
cm$^{-2}$ in Fig. 1b. These traces span nearly the entire range of 
the densities in this low-$T$ study, and it is clear from the  
appearance of the Shubnikov--de Haas oscillations and the quantum Hall 
features that homogeneity and uniformity are adequately maintained 
throughout. 

We now turn to our main results. In Fig. 2 we plot the 
resistivity versus $T$ obtained from sample H678D, for 5 values of $n$.
Although not as 
pronounced as in our p-type samples \cite{Hanein1998}, the main features 
of the MIT are 
easily seen. At the lowest density, $n=1.2\cdot 10^{10}$ cm$^{-2}$,
(top curve in Fig. 2a) the sample appears to be insulating, 
with $\rho$ monotonically decreasing with $T$. 
At slightly higher density, $n=1.3\cdot 10^{10}$ cm$^{-2}$, (bottom curve 
in Fig. 2a) 
the resistivity 
is still monotonically decreasing with $T$, but tends to saturate below 
150 mK. We associate the density where the resistivity is $T$-independent  
at our lowest $T$'s with the metal-insulator transition point.
The procedure for identifying the metal-insulator transition point
was described in more details in Ref. \cite{Hanein1998prb} for our  
p-type samples. The critical 
resistivity obtained for the sample in this work is $10.4$ k$\Omega$.

In the next three graphs (Fig. 2b-2d) the sample exhibits metallic 
behavior at low $T$. In Fig. 1b, non-monotonic dependence of 
$\rho$ is observed, with insulating-like $\rho(T)$ at higher $T$ 
changing over to metallic-like behavior at our lower $T$ range. This 
non-monotonic behavior invariably appears near the transition to the insulating phase. 
At the next higher $n$ (Fig. 2c) a pure metallic behavior is seen with a 
pronounced drop 
in $\rho$ at lower $T$'s. Finally, Fig. 2d shows  
a high-$n$ trace where $\rho$ is only weakly dependent on $T$ over our 
entire $T$ range. We wish to emphasize that although the qualitative 
features of these data are very similar to those obtained from 
Si and p-type GaAs-based materials, the magnitude of the features near 
the 
transition are much reduced. 
For instance, the largest percentage drop in $\rho$ 
we observe for our $n$-type GaAs is $<3\%$, to be contrasted with a 
ten-fold drop in the case of the highest mobility Si-MOSFET's, or a 
factor of 3 for our p-type GaAs samples. We note that metallic 
behavior is observed in the sample of this work down to our lowest $T$ and 
at $n\leq 6\cdot 10^{10}$ cm$^{-2}$. No weak-localization 
corrections to $\rho(T)$ could be resolved within our experimental 
accuracy.

An additional similarity between the n-type and 
the p-type metallic conductivities is the   
dependence of $\rho$ on $T$. In both cases it appears to obey the form :
\begin{equation}
\rho(T)=\rho_0+\rho_1\exp(-\frac{T_0}{T})
\label{PudFit}
\end{equation} 
This relation was previously observed in n-type Si-MOSFET's 
\cite{Pudalov1997}.
In Fig. 2c we include a fit (dashed line) of the data to Eq. \ref{PudFit}.
While the quality of the fit is certainly good, we emphasize that 
the range of the variations in $\rho(T)$ is much too small, 
and is insufficient to determine the true functional form of 
$\rho(T)$. The $T_{0}$ obtained from the fit is $0.26$ K.

The addition of the n-type GaAs-based 2D system to the list of materials 
that exhibit the MIT serves as a convenient opportunity to consider 
some general aspects of the transition and their bearing on various 
material-dependent parameters. 
In particular we address the question 
of the univerality of the resistivity value at the transition. 
In Fig. 3 we plot a compilation of the critical $\rho$ value, 
$\rho_{c}$, versus critical density, $n_{c}$, obtained from our samples 
together with similar published results of several other groups.


In contrast with studies of the quantum Hall to insulator transitions 
at high magnetic field, where critical resistivities obtained from 
many samples were within 30\% of $h/e^{2}$ \cite{DShahar95,Wong95}, in 
the MIT-in-2D case the 
scatter of the data points is significantly larger, even when  
the somewhat anomalous results from n-type Si/SiGe samples 
of Refs. \cite{Ismail} are discarded. 
Further, systematic variations in $\rho_{c}$ exist, as pointed out in 
Ref. \cite{Pud98}. However, one can not ignore the impression 
created by the data that 
most transition values are scattered around $\rho_{c}=1$. The question 
of whether this 
is a coincidental observation, or the $\rho_{c}=1$ point has some 
special significance, awaits further experimental as well as theoretical 
studies. 

Summarizing, we demonstrated the existance of a MIT in 2D, n-type, GaAs 
material. To date, it is the highest mobility sample that exhibit this 
transition.

This research was supported by the Israel Science Foundation founded 
by The Academy of Science and Humanities, the
NSF and by a grant from the Israeli 
Ministry of Science and The Arts (YH).

\begin{figure}  
\caption{$\rho_{xx}$ and $\rho_{xy}$ versus magnetic field, at low-$T$, 
for two densities.
(a) Low density traces, $n=1.3\cdot 10^{10}$ cm$^{-2}$ and $\mu$=47,000 
cm$^{2}$/Vsec
(b) High density traces, $n=6\cdot 10^{10}$	cm$^{-2}$ and 
$\mu$=890,000 cm$^{2}$/Vsec.}
\end{figure}
 
\begin{figure}  
\caption{$\rho_{xx}$ versus $T$ for 5 different densities.
$\rho$ of the 2DES changes from insulating to metallic as the density 
is increased.
(a) Trace I: $n=1.2\cdot 10^{10}$ cm$^{-2}$ 
($r_{s}=4.8$), and trace II: $n=1.3\cdot 
10^{10}$ cm$^{-2}$ ($r_{s}=4.6$).
(b) $n=1.9\cdot 10^{10}$ cm$^{-2}$ ($r_{s}=3.8$).
(c) $n=2.5\cdot 10^{10}$ cm$^{-2}$ 
($r_{s}=3.3$), and a fit to Eq. \ref{PudFit}, the fit is denoted by a 
dashed line.
(d) $n=4.2\cdot 10^{10}$ cm$^{-2}$ ($r_{s}=2.5$).}
\end{figure}
 
\begin{figure}  
\caption{Resistivity versus density at the 
transition for various samples from different published studies. 
The full markers denote Si-based materials which are  
characterized by large critical densities $n_{c}>7\cdot 10^{10}$ 
cm$^{-2}$. The open markers denote 
GaAs-based materials,
characterized by smaller critical densities, $n_{c}<7\cdot 10^{10}$ 
cm$^{-2}$. }
\end{figure}

\end{document}